\documentclass[9pt,conference]{IEEEtran}
\usepackage[T1]{fontenc}

\usepackage[preprint]{waspaa25} 

\usepackage{bm} 
\usepackage{booktabs}
\usepackage{siunitx}
\usepackage{tabularx}
\usepackage{makecell}
\usepackage{color}
\usepackage{subcaption}
\usepackage{tablefootnote}
\usepackage{multirow}
\usepackage{bm}

\usepackage{bm} 
\usepackage{booktabs}
\usepackage{siunitx}
\usepackage{color}
\usepackage{subcaption}
\usepackage{tablefootnote}

\newcolumntype{Y}{>{\centering\arraybackslash}X}

\title{Treble10: A high-quality dataset for far-field speech recognition, dereverberation, and enhancement
}

\name{Sarabeth S. Mullins$^{1*}$, Georg G\"otz$^{1}$, Eric Bezzam$^{2}$, Steven Zheng$^{2}$, Daniel Gert Nielsen$^{1}$
}
\address{$^{1}$Treble Technologies, Reykjavík, Iceland\\ $^{2}$Hugging Face, Paris, France\\ $^{*}$Correspondence: ssm@treble.tech}

\begin{document}

\maketitle

\begin{abstract}
Accurate far-field speech datasets are critical for tasks such as automatic speech recognition (ASR), dereverberation, speech enhancement, and source separation. However, current datasets are limited by the trade-off between acoustic realism and scalability. Measured corpora provide faithful physics but are expensive, low-coverage, and rarely include paired clean and reverberant data. In contrast, most simulation-based datasets rely on simplified geometrical acoustics, thus failing to reproduce key physical phenomena like diffraction, scattering, and interference that govern sound propagation in complex environments. We introduce Treble10, a large-scale, physically accurate room-acoustic dataset. Treble10 contains over 3000 broadband room impulse responses~(RIRs) simulated in 10 fully furnished real-world rooms, using a hybrid simulation paradigm implemented in the Treble SDK that combines a wave-based and geometrical acoustics solver. The dataset provides six complementary subsets, spanning mono, 8th-order Ambisonics, and 6-channel device RIRs, as well as pre-convolved reverberant speech scenes paired with LibriSpeech utterances. All signals are simulated at \SI{32}{\kilo\Hz}, accurately modelling low-frequency wave effects and high-frequency reflections. Treble10 bridges the realism gap between measurement and simulation, enabling reproducible, physically grounded evaluation and large-scale data augmentation for far-field speech tasks. The dataset is openly available via the Hugging Face Hub, and is intended as both a benchmark and a template for next-generation simulation-driven audio research.
\end{abstract}

\section{Introduction}
Accurate room-acoustic data is the foundation for far-field speech recognition, dereverberation, speech enhancement, or source separation. Yet, most existing datasets are limited either in scale or realism. Measured corpora, such as the BUT ReverbDB \cite{Szöke2019ButReverbDatasetPaper} or the CHIME3 challenge dataset \cite{Barker2015ThirdChime} capture acoustic conditions of the measured scenes reliably, but only coarsely cover selected areas of the rooms. For example, BUT ReverbDB contains around 1400 measured room impulse responses (RIRs) from 9 rooms, recorded with a sound source and a few dozen microphones inside spatially constrained areas. Expanding or replicating these datasets is extremely time-consuming and expensive. The CHiME datasets offer much larger amounts of noisy, reverberant speech but usually do not provide the underlying RIRs. The lack of matching RIR data limits the usability for tasks like dereverberation, where both original (aka ``dry'') and reverberant versions of the same utterance are required. Additionally, systematic data generation and controlled ablations are not possible in such cases.

The Treble10 dataset bridges this gap by combining physical accuracy with the scalability of advanced simulation. Using the Treble SDK’s hybrid wave-based and geometrical-acoustics engine, we model sound propagation in 10 realistic, fully furnished rooms. In contrast to other simulation tools, which typically rely on simplified geometrical acoustics modelling, our hybrid approach models physical effects such as scattering, diffraction, interference, and the resulting modal behaviour. Each room of the Treble10 dataset is densely sampled across receiver grids at multiple heights, resulting in over 3000 distinct transfer paths per subset. The dataset includes 6 subsets: mono, 8th-order Ambisonics, and 6-channel device RIRs, along with corresponding reverberant speech signals from the LibriSpeech~\cite{Panayotov2015Librispeech} test set (test.clean and test.other). All responses are broadband (\SI{32}{\kilo\Hz} sampling rate), accurately modelling both low-frequency wave behaviour and high-frequency reflections.

The dataset is openly available via the Hugging Face Hub:
\begin{itemize}
    \item[] \url{https://huggingface.co/collections/treble-technologies/treble10}
\end{itemize}
Code examples that demonstrate how to use the dataset can be found in the respective dataset cards on the Hugging Face Hub.

\section{Motivation}
\subsection{Near-field vs. far-field algorithm development}
\begin{figure}[tb]
\centering
    \includegraphics[width=0.5\textwidth]{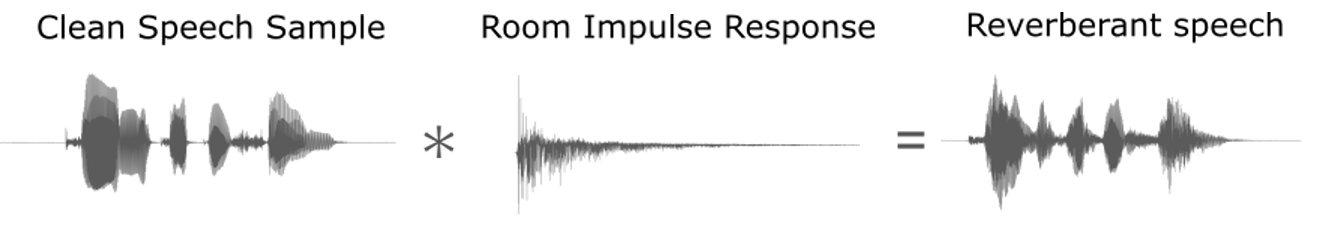}
    \caption{The transformation of a clean into a reverberant audio signal via convolution with a simulated Room Impulse Response (RIR). The clean speech (left) is convolved with the RIR (centre). This operation yields augmented clean speech (right) that incorporates the acoustic characteristics of that particular room and source-receiver combination.}
    \label{fig:rir_conv}
\end{figure}
Which use cases benefit from the Treble10 dataset? Let us explain this briefly with an example scenario. Consider the difference between far-field and near-field automatic speech recognition (ASR).

\begin{figure*}[t]
    \includegraphics[width=\textwidth]{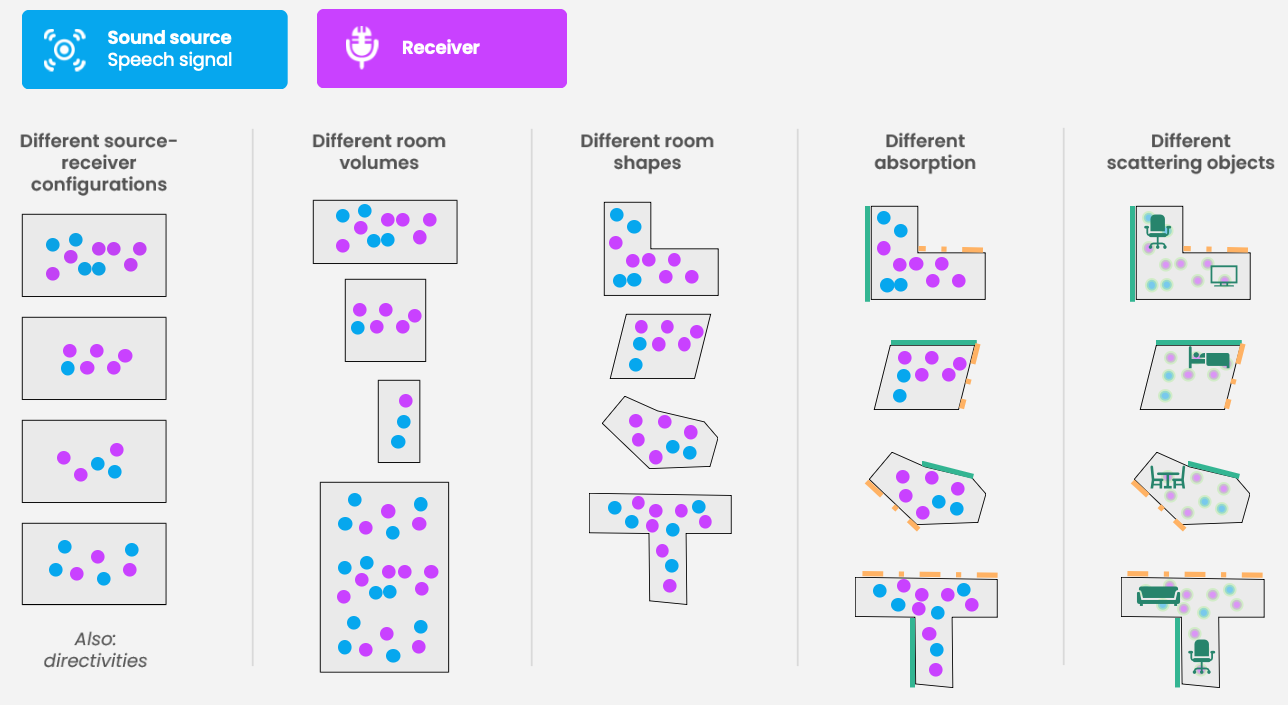}
    \caption{Disentangling the different degrees of freedom of room-acoustic datasets.}
    \label{fig:disentanglement}
\end{figure*}

In near-field ASR, a user speaks directly into a smartphone or headset, and the captured speech signal is relatively clean. The microphone is close to the mouth, so the direct sound dominates while reverberation and background noise are comparatively weak. In these conditions, ASR models may perform well even with limited room-acoustic diversity in the training data. However, in far-field ASR, as in smart speakers or conference-room devices, the microphone may be located several meters from the talker. The speech signal reaching the microphone is a complex mixture of direct sound, reverberation, and background noise, making the recognition task substantially more challenging.

The difference between near-field and far-field conditions is not just a matter of distance, but also a matter of physics. In far-field setups, sound interacts heavily with the room: it reflects off walls, diffracts around furniture, and decays over time. RIRs comprise all of these effects and encode how sound propagates from the source to the receiver. By convolving a dry audio signal with an RIR, as illustrated in Fig. \ref{fig:rir_conv}, we can simulate reverberant speech for a specific room configuration, replicating the far-field scenario.

For far-field ASR systems to be robust, they must be trained on data that accurately represents the complex far-field behaviour \cite{Kim2017SimulatedFarField}. Similarly, the performance of far-field ASR systems can only be reliably determined when evaluating them on data that is accurate enough to model sound propagation in complex environments.

This example illustrates the need for accurate RIR datasets in far-field ASR. Other speech-related tasks, such as speech enhancement, dereverberation and source separation, in real-world conditions benefit analogously from high-quality training data.

\subsection{Importance of domain data}
Figure \ref{fig:disentanglement} shows that room-acoustic data exhibits many degrees of freedom. To make algorithms perform well in a broad range of realistic conditions, we want the underlying training data to cover different source-receiver configurations, different source and receiver directivities or devices, different room volumes, different room shapes, different wall absorption properties, and different geometric details like scattering objects.

However, when developing audio algorithms or training machine learning models for acoustic tasks, a central question inevitably arises: Where can we obtain sufficiently large and high-quality room-acoustic datasets that cover all outlined dataset dimensions? Room-acoustic measurements capture the physical sound pressure within a space at a specific moment in time, and therefore, faithfully represent the actual acoustic conditions of that environment at the measurement time. Unfortunately, conducting such measurements is both labour-intensive and time-consuming.

Even with recent advances in automated room-acoustic measurements \cite{Goetz2021ARTSRAM,Stolz2023AutonomousRoboticPlatformSRIR,Goetz2023AutonomousRoomAcousticMeasurementsRRTGaussianProc}, large-scale, measurement-based dataset acquisition that sufficiently covers the aforementioned dataset dimensions remains impractical. Such large measurement campaigns would not only require tremendous amounts of manual labour, but also quickly reach practical limits regarding scalability. For example, a research team may have physical access to ten or even a hundred different rooms, but scaling such a campaign to the order of ten or even a hundred thousand rooms is infeasible. Setting up device-specific multi-channel datasets further multiplies the measurement effort, as new measurements have to be conducted for every device configuration.

\subsection{How simulations can help}
Room-acoustic simulations are a scalable tool for generating large amounts of synthetic data under controlled, reproducible, and diverse acoustic conditions. This makes them a powerful alternative to physical measurements for training and evaluating audio machine learning models. A wide range of simulation methods exist, from geometrical acoustics techniques like the image-source method \cite{Allen1979ImageMethodForEfficientlySimulatingSmallRoomAcoustics,Borish1984ExtensionOfTheImageModelToArbitraryPolyhedra} and ray tracing \cite{Krokstad1968CalculatingTheAcousticalRoomResponseByTheUseOfRayTracing,Krokstad1983Fifteenyearsexperiencewithcomputerizedraytracing}, to more advanced, wave-based approaches that capture the underlying physics of sound propagation in greater detail~\cite{BottelDooren1995FDTDSimulationRoomAcoustic,Wang2019DiscontinuousGalerkinRoomAcoustics,Pind2020TimeDomainSimExtendedReactingPorousDGFEM}.

Reproducing real-world acoustics with simulations requires accurate modelling of key acoustic phenomena that shape how sound interacts with the environment, including:
\begin{itemize}
\item \textbf{Reflection and scattering:} how sound waves reflect on surfaces and diffuse from rough or textured materials;
\item \textbf{Diffraction:} how sound bends around obstacles such as furniture, people, or architectural features;
\item \textbf{Absorption:} how different materials attenuate sound energy in a frequency-dependent manner; and
\item \textbf{Reverberation:} the complex temporal decay of reflections that gives each room its distinctive acoustic character.
\item \textbf{Sound source and receiver directivity:} how strongly a sound source radiates in different directions and how sensitive a device or microphone is to sounds coming from different directions.
\end{itemize}

\begin{figure*}[t]
\centering
    \includegraphics[width=\textwidth]{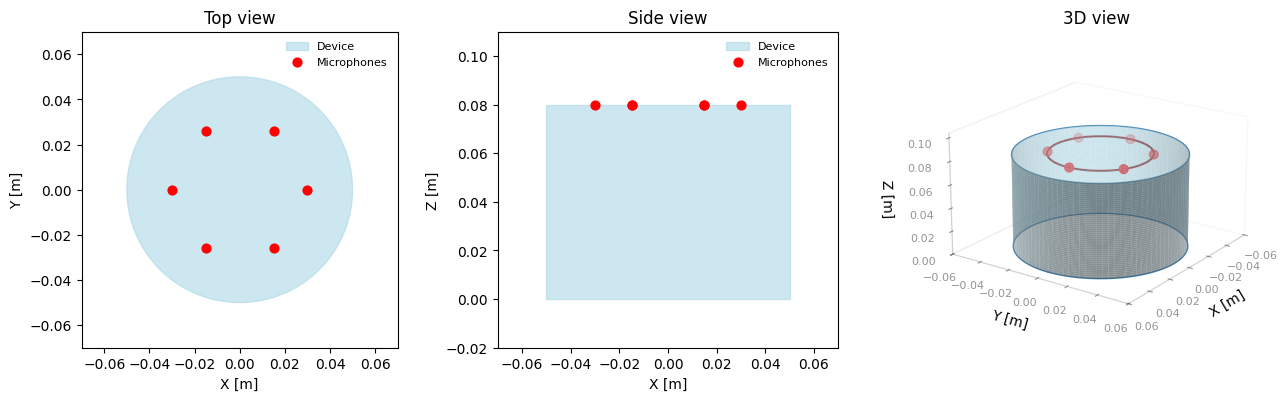}
    \caption{Sketch of the multi channel device included in the dataset. The device consists of $6$ microphones evenly spaced with a radius of \SI{3}{\centi\meter}. The coordinates of the microphones can be found in the metadata for the $6$ch split and also in the dataset card.}
    \label{fig:device}
\end{figure*}

However, many existing data augmentation pipelines still rely on simplified simulation techniques, such as basic image-source models and ray tracing. Although these approaches can be effective for simple environments, they often fail to capture the full physical complexity of real-world acoustics, particularly in spaces with irregular geometries, frequency-dependent materials, scattering objects, or directive source and receivers. Consequently,  machine learning algorithms trained with such simplified simulation data exhibit inferior model performance compared to algorithms trained with more realistic data \cite{Bezzam2020MoreRealisticRoomSimulationForKeywordSpotting,Srivastava2023HowToVirtuallyTrain,Arakawa2024QuantifyingSimulatorBasedDataAugmentation,Guso2025MBRIRs}.

\section{The Treble10 dataset}
The Treble10 dataset contains high fidelity room-acoustic simulations from 10 different furnished rooms, as summarized in Table \ref{tbl:Rooms}. It contains six subsets:
\begin{enumerate}
    \item \textbf{Treble10-RIR-mono:} This subset contains mono RIRs. In each room, RIRs are available between 5 sound sources and several receivers. The receivers are placed along horizontal receiver grids with \SI{0.5}{\meter} resolution at three heights (\SI{0.5}{\meter}, \SI{1.0}{\meter}, \SI{1.5}{\meter}). The validity of all source and receiver positions is checked to ensure that none of them intersects with the room geometry or furniture.
    \item \textbf{Treble10-RIR-HOA8:} This subset contains 8th-order Ambisonics RIRs. The sound source and receiver positions are identical to the RIR-mono subset.
    \item \textbf{Treble10-RIR-6ch:} For this subset, a $6$-channel cylindrical device (see Fig. \ref{fig:device}) is placed at the receiver positions from the RIR-mono subset. RIRs are then acquired between the $5$ sound sources from above and each of the $6$ device microphones. In other words, there is a $6$-channel DeviceRIR for each source-receiver combination of the RIR-mono subset. The microphone coordinates along with an example of how to use the multichannel device can be found in the dataset card\footnote{\url{https://huggingface.co/datasets/treble-technologies/Treble10-RIR}}.
    \item \textbf{Treble10-Speech-mono:} Each RIR from the RIR-mono subset is convolved with a speech file from the LibriSpeech test set~(test.clean and test.other).
    \item \textbf{Treble10-Speech-HOA8:} Each Ambisonics RIR from the RIR-HOA8 subset is convolved with a speech file from the LibriSpeech test set~(test.clean and test.other).
    \item \textbf{Treble10-Speech-6ch:} Each DeviceRIR from the RIR-$6$ch subset is convolved with a speech file from the LibriSpeech test set~(test.clean and test.other).
\end{enumerate}

All RIRs (mono/HOA/device) were simulated with the Treble SDK\footnote{\url{https://www.treble.tech/software-development-kit}}, and more details about the tool can be found in Sec. \ref{sec:TrebleSDK}. We use a hybrid simulation paradigm that combines a numerical wave-based solver (discontinuous Galerkin method \cite{Wang2019DiscontinuousGalerkinRoomAcoustics,Pind2020TimeDomainSimExtendedReactingPorousDGFEM}, DGM) at low- to mid-range frequencies with geometrical acoustics (GA) simulations \cite{Savioja2015OverviewGA} at high frequencies. For the Treble10 dataset, the transition frequency between the wave-based and the GA simulation is set at \SI{5}{\kilo\Hz}. The resulting hybrid RIRs are broadband signals with a sampling rate of \SI{32}{\kilo\Hz}, covering the entire frequency range of the signal and containing audio content up to \SI{16}{\kilo\Hz}.

\begin{table}[tb]
\centering
\sisetup{
    reset-text-series = false, 
    text-series-to-math = true, 
    mode=text,
    tight-spacing=true,
    round-mode=places,
    round-precision=2,
    table-format=2.2,
    table-number-alignment=center,
    detect-weight=true,
    detect-family=true
}
\caption{Rooms covered by the Treble10 dataset.}
\label{tbl:Rooms}
\begin{tabular}{lcc}
\toprule
\textbf{Room}                       & \begin{tabular}[c]{@{}c@{}}\textbf{Volume}\\ \textbf{(\si{\meter^3})}\end{tabular} & \begin{tabular}[c]{@{}c@{}}\textbf{Reverberation time}\tablefootnote{Averaged over all source-receiver configurations and octave bands.}\\ $\bm{T_{30}}$ \textbf{(\si{\second})}\end{tabular} \vspace{0.25em}\\ \hline
Bathroom 1                 & 15.42       & 0.58                                                                                            \\
Bathroom 2                 & 18.42       & 0.77                                                                                            \\
Bedroom 1                  & 15.6        & 0.43                                                                                            \\
Bedroom 2                  & 17.65       & 0.22                                                                                            \\
Living room with hallway 1 & 38.66       & 0.62                                                                                            \\
Living room with hallway 2 & 46.08       & 0.62                                                                                            \\
Living room 1              & 40.91       & 0.37                                                                                            \\
Living room 2              & 43.16       & 0.87                                                                                            \\
Meeting room 1             & 13.83       & 0.38                                                                                            \\
Meeting room 2             & 23.97       & 0.19                                                                                            \\ \bottomrule
\end{tabular}
\end{table}

A small subset of simulations from the same rooms has previously been released as part of the Generative Data Augmentation (GenDA) challenge at ICASSP 2025 \cite{Lin2025GenDARAChallenge}. However, the Treble10 dataset differs from the GenDA dataset in three fundamental aspects:
\begin{enumerate}
\item The Treble10 dataset contains broadband RIRs from a hybrid simulation paradigm (wave-based below \SI{5}{\kilo\Hz}, GA above \SI{5}{\kilo\Hz}), covering the entire frequency range of a \SI{32}{\kilo\Hz} signal. In contrast to the GenDA subset, which only contained the wave-based portion, the Treble10 dataset therefore more than doubles the usable frequency range.
\item The Treble10 dataset consists of $6$ subsets in total. While three of those subsets contain RIRs (mono, $8$th-order Ambisonics, $6$-channel device), the other three contain pre-convolved scenes in identical channel formats. The GenDA subset was limited to mono and 8th-order Ambisonics RIRs, and no pre-convolved scenes were provided.
\item With Treble10, we publish the entire dataset, containing approximately $3100$ source-receiver configurations. The GenDA subset only contained a small fraction of approximately $60$ randomly selected source-receiver configurations.
\end{enumerate}

\section{Treble Technologies and the Treble SDK}
\label{sec:TrebleSDK}
At Treble Technologies, we believe that advancing audio machine learning requires revisiting the underlying physics of sound propagation. A physically accurate simulation paradigm is crucial, especially when modelling the complex behaviour of sound in realistic rooms and with multi-channel devices. Tools like the Treble SDK bridge the gap between high scalability and high accuracy when working with room-acoustic simulations.

The Treble SDK offers a flexible, Python-based framework powered by an advanced acoustic simulation engine. It enables engineers and researchers to:
\begin{itemize}
\item Simulate complex acoustic environments: Move beyond simple room models to capture detailed geometries and devices. Model wave effects that are crucial phenomena in real-world sound propagation, such as diffraction, scattering, interference, and the resulting modal behaviour.
\item Generate multi-channel RIRs at scale: Create large datasets of physically accurate, high-fidelity RIRs, including precise simulations for custom microphone arrays embedded within device structures.
\item Control acoustic parameters with precision: Adjust room size, material properties, source and receiver positions (including array layouts), and object placements to produce diverse and well-controlled acoustic conditions.
\end{itemize}

\bibliographystyle{IEEEtran}
\bibliography{refs25}

\end{document}